\documentclass[sigconf, authorversion]{acmart}
\usepackage[acronym]{glossaries}
\usepackage{xcolor}
\makeglossaries

\newacronym{sbr}{SBR}{Session-Based Recommendation}
\newacronym{gnn}{GNN}{Graph Neural Networks}
\newacronym{nlp}{NLP}{Natural Language Processing}
\newacronym{mlm}{MLM}{Masked Language Modeling}
\newacronym{clm}{CLM}{Causal Language Modeling}
\newacronym{smm}{SMM}{Sequential Masked Modeling}
\newacronym{llm}{LLM}{Large Language Model}
\newacronym{rnn}{RNN}{Recurrent Neural Network}
\newacronym{rope}{RoPE}{Rotary Position Embedding}
\newacronym{cope}{CoPE}{Contextual Position Encoding}

\AtBeginDocument{%
  }

\renewcommand\footnotetextcopyrightpermission[1]{} 

\begin{document}

\title{Optimizing Encoder-Only Transformers for Session-Based Recommendation Systems}

\author{Anis Redjdal}
\authornote{Graduate student conducting research in the domain. This work forms part of the research toward the completion of a Master’s thesis.}
\email{anis.redjdal.poly@gmail.com}
\orcid{1234-5678-9012}
\affiliation{%
  \institution{Polytechnique Montreal}
  \city{Montreal}
  \state{Quebec}
  \country{Canada}
}

\author{Luis Pinto}
\email{luis.pinto1@airliquide.com}
\orcid{1234-5678-9012}
\affiliation{%
  \institution{Air Liquide}
  \city{Montreal}
  \state{Quebec}
  \country{Canada}
}

\author{Michel Desmarais}
\email{michel.desmarais@polymtl.ca}
\orcid{1234-5678-9012}
\affiliation{%
  \institution{Polytechnique Montreal}
  \city{Montreal}
  \state{Quebec}
  \country{Canada}
}

\begin{abstract}

Session-based recommendation is the task of predicting the next item a user will interact with, often without access to historical user data. In this work, we introduce Sequential Masked Modeling, a novel approach for encoder-only transformer architectures to tackle the challenges of single-session recommendation. Our method combines data augmentation through window sliding with a unique penultimate token masking strategy to capture sequential dependencies more effectively. By enhancing how transformers handle session data, Sequential Masked Modeling significantly improves next-item prediction performance.

We evaluate our approach on three widely-used datasets, Yoochoose 1/64, Diginetica, and Tmall, comparing it to state-of-the-art single-session, cross-session, and multi-relation approaches. The results demonstrate that our Transformer-SMM models consistently outperform all models that rely on the same amount of information, while even rivaling methods that have access to more extensive user history. This study highlights the potential of encoder-only transformers in session-based recommendation and opens the door for further improvements.

\end{abstract}



\maketitle

\section{Introduction}
Traditional recommendation systems predominantly rely on a user's historical interactions and preferences \cite{lu2015recommender}. However, when user identities are partially known or entirely anonymous, recommendations must be generated based solely on the actions taken within a single session. These session-based scenarios challenge traditional models, which depend heavily on extensive user-item interaction history to provide accurate recommendations.

\gls{sbr} models tackle this challenge by analyzing user behavior within a single session to predict future actions and make relevant recommendations. Each session is treated as an independent, ordered sequence of consecutive interactions with items, regardless of whether the same user appears in multiple sessions, focusing solely on the interactions within a specific context. Accurately modeling these sessions is crucial for improving recommendation relevance and providing a personalized, engaging user experience. For example, on an e-commerce website, a session might begin when the user logs in and conclude when they leave the site, capturing the sequence of items they viewed in chronological order.

Considering that the task of \gls{sbr} is to predict the next item that might interest a user, the task closely resembles the challenges associated with predicting the next word in \gls{nlp}. This parallel is a natural motivation to explore the application of transformer models \cite{vaswani2017attention}. Just like in \gls{nlp}, items are represented as tokens and processed as a sequence. 

In this work, we introduce \gls{smm}, a novel approach for improving session-based recommendation using encoder-only transformer architectures. Inspired by the success of transformers in NLP, our method adapts these models to \gls{sbr} by employing data augmentation through window sliding and a new masking strategy. Additionally, we optimize the transformer architectures to further enhance performance. We evaluate our proposed \gls{smm} method on three benchmark datasets: Yoochoose 1/64, Diginetica, and Tmall. Our Transformer-SMM models were evaluated against state-of-the-art models, demonstrating clear improvements when working with the same level of information. Even in comparison to models utilizing more user data, our approach remained competitive in terms of precision and ranking metrics. In the following sections, we describe our method, the optimization techniques applied, and evaluate our approach against existing benchmarks.

\begin{figure}
\centering
\includegraphics[width=3in]{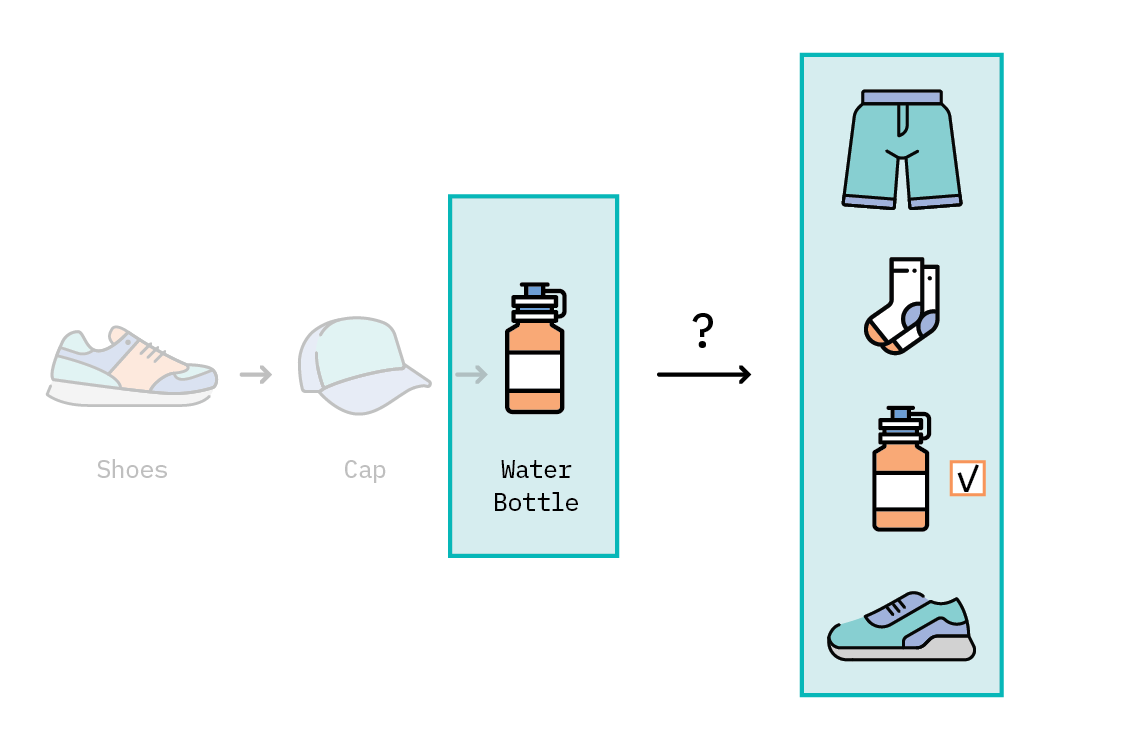}
\caption{Next-click prediction task \cite{cloudera2021session}.}
\label{fig:next_click_pred}
\end{figure}

\section{Related work} 

Early research in session-based recommendations primarily explored sequential models, with the Markov Chain (MC) model being one of the first to address this domain \cite{Dias2005}. However, MC often struggles with capturing long-range dependencies in the user's behavior, as it tends to compress information from earlier interactions. As a result, its effectiveness diminishes as the sequence length increases.

Hidasi et al. \cite{hidasi2015session} introduced a \gls{rnn}-based method, which was later enhanced by incorporating data augmentation and accounting for temporal shifts in user behavior, as explored by \cite{velivckovic2017graph}. Further developments in 2016 introduced the concept of integrating dwell time, the duration of item views, into session-based recommendations \cite{hidasi2016incorporating}. This enhancement allowed \gls{rnn} models to more accurately capture user interest and engagement, thereby boosting their predictive performance. Building on this, NARM \cite{li2017neural} introduced a dual-approach \gls{rnn} recommendation system that captures both the sequential patterns of user behavior and their primary intentions. Despite the inherent challenges associated with \gls{rnn}s, such as the difficulty in representing users with limited data or the complexity of modeling distant item transitions within a sequence, researchers have gradually improved this methodology over time, though the associated challenges remain significant.

To address these limitations, the introduction of \gls{gnn} in 2019 marked a major advancement in the field of \gls{sbr}. Session-based \gls{gnn}s \cite{xu2019graph, huang2021graph, zhang2023graph} are specifically designed to generate representations within graphs of item interactions, making them particularly effective at capturing and representing the intricate patterns in these sequences. Since the introduction of SR-GNN \cite{xu2019graph}, graph-based models have gained significant recognition in \gls{sbr}, consistently delivering superior performance. Over time, models like CARES \cite{zhang2023context}, STAR \cite{ahn2023star}, CoTREC \cite{xia2021self}, and GCE-GNN \cite{wang2020global} have further refined this approach, positioning \gls{gnn}s at the current state-of-the-art in \gls{sbr}.

Transformers have also achieved significant success in the SBR domain. In the RecSys Challenge 2022 \cite{RecSysChallenge2022}, a team employing transformer architectures secured 2nd place among 300 participants \cite{lu2022session}. Furthermore, Zhang et al. \cite{zhang2023graph} introduced a novel approach that integrated a \gls{gnn} with a transformer, using an attention mechanism to process sequences in graph form. This hybrid model, based on the BERT architecture \cite{devlin2018bert}, primarily utilizes an encoder to enhance session data representations. The fusion of graph-based models and transformer architectures marks a significant step forward in \gls{sbr}, combining the strengths of both techniques.

Building on the success of transformers in \gls{sbr}, \textit{Meta}, in collaboration with \textit{Nvidia}, developed a dedicated library that provides the most popular transformer architecture for session-based recommendation \cite{de2021transformers4rec}. While their reported performance on benchmarks falls short of \gls{gnn}-based models, this library offers a useful solution by streamlining session data preprocessing and transformer language model training into a single pipeline. 



\section{Preliminaries}

\subsection{Problem Statement}

First, we formally define the task of next-click prediction in session-based recommendations. Let $V = \{v_1, v_2, \dots, v_m\}$ represent the set of all items, where $m$ denotes the number of items in $V$. Assume that all sessions are represented as $U = \{S_1, S_2, \dots, S_n\}$, where $n$ is the total number of sessions. Each session $S_{\tau} \in U$, denoted as $S_{\tau} = \{v_{\tau 1}, v_{\tau 2}, \dots, v_{\tau t}\}$, consists of a sequence of interactions in chronological order, where $v_{\tau t}$ represents the item interacted with by the user at the $t$-th timestamp in session $S_{\tau}$, and the length of $S_{\tau}$ is $t$. 

The goal of \gls{sbr} is to recommend the next item from $V$ that the user is most likely to interact with, given the current session $S_{\tau}$. Specifically, the item interacted with at the $(t + 1)$-th timestamp is referred to as the target item or ground truth item of the session. Thus, a session and its target item pair can be expressed as $([v_1, v_2, \dots, v_t], v_{t+1})$.

\subsection{Language Models}

Motivated by the success of transformers in \gls{nlp}, where they perform exceptionally at capturing sequential dependencies and contextual information, we apply these models to session-based recommendation. With their ability to model long-range dependencies through the attention mechanism, transformers have proven highly effective in tasks like next-word prediction, and this same architecture can be adapted to predict the next item in a user's browsing session. While the data and context differ from NLP, the underlying principle of predicting the next interaction based on past behavior remains similar, as shown in Figure \ref{fig:next_click_pred}. For the experiments and evaluations conducted in this work, we focused on three popular transformer models: two encoder-only models, BERT \cite{devlin2018bert} and DeBERTa \cite{he2020deberta}, and one decoder-only model, GPT \cite{lee2020patent}.

These language models are trained with a technique called masking. This approach helps the model learn contextual relationships and dependencies between tokens by hiding parts of the input sequence and requiring the model to predict the missing elements. This method is particularly effective for capturing patterns and structure within sequences, making it essential for a variety of tasks, including session-based recommendations. Two masking techniques are commonly used: \gls{mlm} and \gls{clm}.

\subsubsection{Masked Language Modeling}

This is a training technique where a random selection of tokens within a sequence is masked, and the model is trained to predict the masked tokens based on the surrounding unmasked context. By leveraging both preceding and following tokens, this approach enables the model to learn bidirectional contextual representations, enhancing its ability to capture the full context of the input data. The models that employ this masking approach are BERT and DeBERTa. The corresponding loss function is as follows:

\[
\mathcal{L}_{\text{MLM}} = -\sum_{i \in \mathcal{M}} \log P(x_i | x_{\setminus i})
\]

where $\mathcal{M}$ is the set of masked positions, $x_i$ is the token at position $i$, and $x_{\setminus i}$ represents all tokens except $x_i$.

\subsubsection{Causal Language Modeling}

Also known as autoregressive modeling, trains the model to predict each token in a sequence using only the preceding context. Since each prediction is based on the tokens preceding the target token, the model is limited to unidirectional attention. The model that we used for this masking technique is GPT, trained with the following loss function:

\[
\mathcal{L}_{\text{\gls{clm}}} = -\sum_{i=1}^{n} \log P(x_i | x_1, x_2, \ldots, x_{i-1})
\]

where \(x_i\) represents the \(i\)-th token in the sequence, and \(n\) is the total number of tokens. The model estimates the likelihood of the \(i\)-th token given all previous tokens \(x_1, x_2, \ldots, x_{i-1}\). This approach enables the model to generate coherent and contextually relevant sequences, making it particularly useful for tasks like text generation. However, the unidirectional nature of this model means it cannot consider future tokens during training (when computing scores for each token), which may limit its performance in tasks that require understanding both preceding and following contexts.

\section{The proposed method}

In this section, we introduce Sequential Masked Modelling (\gls{smm}), a new approach designed to improve next-click prediction in session-based recommender systems. This method employs a novel masking technique during training to enhance performance.

The \gls{smm} approach consists of two key principles: data augmentation with window sliding, and masking the penultimate token of the augmented sequences. This masking approach is specifically applied to encoder-only transformer models, as it leverages their bidirectional attention mechanism.

\subsection{Data Augmentation with Window Sliding}

In transformer models, the input sequence length is constrained by a parameter known as \textit{max\_len}, which controls the maximum number of tokens the model can process in a single pass. Due to computational limits and the statistical properties of the input data, sequences that exceed this length must be truncated or split into smaller chunks. In many cases, shrinking the window to fit within this constraint is acceptable without compromising the model’s ability to learn effectively.

To address this, we implement a sliding window technique to divide sequences into fixed-size chunks based on the \textit{max\_len} parameter. This ensures that the model can handle variable-length sequences while adhering to computational limits. Each session is represented as a sequence of items with a minimum length of two\footnote{A minimum length of 2 is necessary because, during training, at least one item must remain unmasked for prediction.}. For sequences longer than two items, subsequences of length \(n - 1\) are generated by progressively removing the last item at each step, where \textit{n} is the total number of items in the sequence.

The sliding window always starts at the end of the sequence and moves leftward, generating subsequences of length \textit{max\_len}. This allows all relevant parts of the sequence to be included in training and evaluation, regardless of the original sequence length. If the sequence length is shorter than \textit{max\_len}, the entire sequence is used.

Given an initial sequence $s = (x_1, x_2, \ldots, x_n)$, the sliding window moves along the sequence to generate subsequences as follows:

\begin{itemize}
    \item \textbf{Step 1}: The first subsequence is extracted from the end of the initial sequence. If the sequence length is shorter than \textit{max\_len}, the subsequence contains all elements. For example, $s_1 = (x_{n-\textit{max\_len}+1}, \ldots, x_n)$.
    \item \textbf{Step 2}: The window then shifts one step to the left to create the next subsequence. For example, $s_2 = (x_{n-\textit{max\_len}}, \ldots, x_{n-1})$.
    \item \textbf{Step 3}: This process continues, with the window sliding leftward until the window reaches the beginning of the initial sequence. For example, $s_3 = (x_{n-\textit{max\_len}-1}, \ldots, x_{n-2})$, and so on.
\end{itemize}

Through this method, each token in the initial sequence $s$ appears in multiple subsequences, allowing the model to encounter each token in varied training contexts. The general form of each subsequence $s_i$ can be defined as:
\[
s_i = (x_{n-\textit{max\_len}-i+2}, \ldots, x_{n-i+1}) \quad \text{for} \quad i = 0, 1, 2, \ldots, n-\textit{max\_len}
\]

\subsection{Masking the Penultimate Token}

In traditional sequence-based tasks, the last item in a sequence is masked, and the sequence, truncated to the model's maximum input length (\textit{max\_len}), is used for prediction. In our approach, we retain this strategy for the base sequence, where we input the original sequence truncated to \textit{max\_len} and mask the last item as usual.


However, with the introduction of data augmentation through sliding windows, we adopt a different masking strategy for the augmented sequences. Instead of masking the last item, as done in the base sequence, we mask the penultimate token in each augmented subsequence. This method enables the model to learn from both the preceding context and minimal future information, resulting in more varied and informative training examples. Specifically, for each input sequence $s = (x_1, x_2, \ldots, x_n)$, the sliding window generates subsequences of length \textit{max\_len}, where the penultimate token $x_{i-1}$ is masked. The loss is then computed based on the model's ability to predict the masked penultimate token using the surrounding context.
\[
\mathcal{L}_{\text{\gls{smm}}} = -\log P(x_{i-1} | x_1, x_2, \ldots, x_{i-2}, x_i)
\]

This masking strategy takes advantage of the bidirectional attention in encoder-only transformer models, such as BERT, allowing the model to leverage both the preceding context and the limited right-hand context (provided by the next element in the sequence). Although real-world next-click prediction tasks lack future context, masking the penultimate token proves useful in scenarios where item order is flexible, such as in e-commerce, where item proximity is often more important than strict order.

By combining data augmentation with this penultimate token masking strategy, we ensure that each token in the sequence is masked at least once during training. This complements the base sequence where the last item is masked, allowing the model to adapt to both the original and augmented contexts. Figure \ref{fig:smm-cute} demonstrates the \gls{smm} training approach using a single base sequence. In this example, the base sequence is augmented into multiple subsequences, and the goal is to predict the masked token \texttt{[M]} in each subsequence. For illustration purposes, the maximum sequence length ($max\_len$) is set to 5.

\begin{figure}[t]
{ \color{blue}
  \begin{itemize}
  \item[]  [\textcolor{red}{5, 6, 8, 4, 98, 56}, 54, 74, 23, 56, [M]]
  \item[]  [\textcolor{red}{5, 6, 8, 4, 98, 56}, 54, 74, 23, [M], 57]
  \item[]  [\textcolor{red}{5, 6, 8, 4, 98}, 56, 54, 74, [M], 56]
  \item[]  [\textcolor{red}{5, 6, 8, 4}, 98, 56, 54, [M], 23]
  \item[]  [\textcolor{red}{5, 6, 8}, 4, 98, 56, [M], 74]
  \item[]  [\textcolor{red}{5, 6}, 8, 4, 98, [M], 54]
  \item[]  [\textcolor{red}{5}, 6, 8, 4, [M], 56]
  \item[]  [5, 6, 8, [M], 98]
  \item[]  [5, 6, [M], 4]
  \item[]  [5, [M], 8]
  \item[]  [[M], 6]
  \end{itemize}

}
\caption{SMM training representation. The \texttt{[M]} token is the masked item the model is trained to predict. \textcolor{red}{Red} tokens are cut from the input due to the maximum sequence length constraint ($max\_len = 5$), and \textcolor{blue}{blue} tokens are the ones visible in the model's input sequence.}
\label{fig:smm-cute}
\end{figure}

\subsection{Hypotheses}\label{hypotheses}

\subsubsection{Advantages of \gls{smm} Over \gls{clm}}\label{hypothese1}

We hypothesize that one of the main advantages of our \gls{smm} approach compared to the Causal Language Modeling used in the GPT architecture is that the attention mechanism changes significantly. In a \gls{clm} model, attention is unidirectional, meaning each token can only \textit{look at} the tokens that precede it. This limits the model’s ability to capture complex relationships between tokens, as the full context of the sequence is not considered.

In contrast, with \gls{smm}, each token in the sequence can look anywhere, thanks to the data augmentation that generates subsequences. This allows the model to learn richer and more contextual representations, as each token has access to the entire sequential context and therefore will have a more accurate attention score.

\subsubsection{Advantages of \gls{smm} Over \gls{mlm}}\label{hypothese2}

One of the key advantages of our \gls{smm} approach over the Masked Language Modeling method lies in its handling of sequences longer than the model’s $max\_len$ parameter. In models trained with \gls{mlm}, when input sequences exceed the $max\_len$, they must be split into smaller segments. This splitting breaks the continuity between tokens that span across sequence boundaries, resulting in the loss of important contextual information and potentially diminishing model performance.

While overlapping subsequences could mitigate this issue, random masking in \gls{mlm} introduces another challenge: a bias toward the tokens in the middle of the sequence. These tokens are more likely to be masked in multiple overlapping segments, whereas tokens near the sequence boundaries are underrepresented. This imbalance can negatively impact the model’s ability to learn representations for tokens at the start or end of a sequence.

Rather than breaking the sequence, \gls{smm} generates subsequences that maintain continuity across the full sequence during training, avoiding these limitations and allowing for a more balanced representation of tokens throughout.

\subsubsection{Masking the penultimate instead of the last}\label{hypothese3}

The \gls{smm} training approach allows the model to take advantage of a small amount of right-side context thanks to the bidirectional attention used in encoder-only transformer architectures, which helps during training. Even though in practice, predicting the next item doesn’t have right-side context (since we don’t have access to what comes next), this method works well in situations where the order of items is not strictly fixed. It's more about the proximity of item pairs in e-commerce data, unlike linguistic data where word order is more rigid. This hypothesis will be experimentally explored and further confirmed in Section \ref{SMM-last-k}, where we compare the effects of masking the penultimate token versus masking other tokens, including the third-to-last, in the sequence.

\section{Optimizing techniques}

In this section, we present several techniques that we introduce to the transformer architectures, which contributed to further improving prediction performance. All the optimizations are implemented across all three transformer models, except for Contextual Positional Encoding, which is only applied to BERT due to compatibility constraints with the attention block. Ablation studies for these optimization techniques, detailing their individual impact, can be found in Appendix \ref{ablation_study}. Furthermore, the optimized BERT architecture can be found in Appendix \ref{bert_optimized}.

\subsection{Weight Tying}

Weight tying \cite{press2016using} technique is used in transformer models to reduce the number of parameters and enhance performance. It involves using the same weights for both the input embeddings and the model’s output layer. In other words, the weight matrices used to encode tokens into embeddings are the same as those used to predict tokens from these representation vectors at the output of our encoder. By sharing the same weight matrix for the embeddings and the output layer, not only is the parameter count reduced, but the model’s generalization is also improved by linking the input and output embeddings.

\subsection{Pre-layer Normalization}

Pre-layer normalization is the second technique implemented across all compatible transformer architectures (GPT, BERT, and DeBERTa). Large Language Models (LLMs), such as Mistral \cite{jiang2023mistral} or Llama \cite{touvron2023llama}, use pre-layer normalization, which applies normalization before the multi-head attention layer and the feed-forward layer. By normalizing before the attention operation, the model can better manage gradients during training, leading to more stable convergence. In our implementation, we use RMSNorm \cite{zhang2019root} instead of layer normalization \cite{ba2016layernormalization}, just like in previously discussed LLM architectures.

\subsection{Contextual Positional Encoding}

We implemented an enhancement to the positional embeddings known as Contextual Positional Encoding (CoPE) \cite{golovneva2024contextualpositionencodinglearning}, which replaces the \gls{rope} method \cite{su2021roformer} commonly used in popular LLMs. Unlike traditional methods that rely on simple token counting, CoPE determines token positions based on contextual information. For compatibility purposes, we applied this technique exclusively to our BERT architecture.

\section{Evaluation and Experiments}

The various techniques we developed were tested through a series of experiments to evaluate their effectiveness. We describe here the methodological details of the experiments.

\subsection{Experimental Datasets}

To evaluate and compare the performance of our transformer models enhanced with our new masking approach and optimization techniques, we selected three publicly available datasets: Yoochoose 1/64 \cite{yoochoose}, Tmall \cite{tmall_dataset}, and Diginetica \cite{diginetica}, which are popular datasets for the session-based recommendation task.

\subsubsection{Data Preprocessing}

The datasets were preprocessed to ensure the results are comparable with various studies in session-based recommendation, particularly those that are considered state-of-the-art and use GNNs \cite{xu2019graph, zhang2023context}. The preprocessing involves three main steps to match the statistics shown in Table \ref{table:dataset_statistics}:

\begin{itemize}
    \item All sessions are ordered chronologically, and the data is split into training and test sets based on session timestamps.
    
    \item We filter out items that appear less than five times or only appear in the test set, as well as sessions of length one.
    
    \item Data augmentation is performed using a sliding window of size \textbf{30} to generate more data samples within a session. For a session [v1, v2, ..., vn], we generate samples ([v1, v2, ..., vn-1], vn), ..., ([v1, v2], v3), ([v1], v2).
\end{itemize}

\begin{table}[h]
\centering
\caption{Dataset Statistics}
\label{table:dataset_statistics}
\begin{tabular}{lrrrr}
\hline
Dataset & \multicolumn{1}{c}{Diginetica} & \multicolumn{1}{c}{Tmall} & \multicolumn{1}{c}{Yoochoose 1/64} \\ \hline
\#Train sessions & 719,470 & 351,268 & 369,859 \\
\#Test sessions & 60,858 & 25,898 & 55,898 \\
\#Items & 43,097 & 40,728 & 16,766 \\
Avg. length & 5.12 & 6.69 & 6.16 \\ \hline
\end{tabular}
\end{table}

\subsection{Implementation}

We implemented all three transformer models (BERT, GPT, and DeBERTa) using the PyTorch library \cite{paszke2019pytorch}. It was not possible to directly import pre-existing models because those models are pre-trained on large text corpora and not on e-commerce items specific to each of our three datasets. Additionally, having access to the internal code of these models was essential to implement recent techniques used by large language models. The code will be available after reviewing.

The three transformer models use the BERT Medium configuration. This configuration consists of 8 encoder layers (decoder for GPT), a hidden size of 512, 8 attention heads, and a total of 41 million parameters \cite{huggingfaceBERTMedium}. We chose this configuration for its balance between computational efficiency and performance. We found that larger configurations often result in overfitting due to the smaller size of e-commerce datasets compared to text corpora, while smaller configurations tend to underfit. The BERT Medium configuration offers a suitable compromise.

\subsection{Metrics}

The evaluations are based on two main metrics:

\subsubsection{Precision@20}

P@20 is defined as the number of relevant recommendations within the top 20 results, divided by 20. It is a commonly used metric to evaluate the performance of recommendation systems. The formula is given by:

\begin{equation}
P@20 = \frac{1}{N} \sum_{i=1}^{N} \frac{\text{relevant predictions at position } \leq 20}{20}
\end{equation}

where \( N \) is the total number of test sessions.

\subsubsection{Mean Reciprocal Rank}

MRR is a metric that evaluates the effectiveness of recommendation systems by considering the position of the first relevant item in the result list. The formula is given by:
\begin{equation}
MRR = \frac{1}{N} \sum_{i=1}^{N} \frac{1}{\text{rank}_i}
\end{equation}

where \( \text{rank}_i \) is the position of the first relevant item in the \( i \)-th session, and \( N \) is the total number of test sessions.

\subsection{Masking Techniques Comparison}

We evaluate and compare the performance of optimized transformer model architectures to determine which masking technique yields the best overall Precision@20 results. The models compared include GPT-CLM, BERT-MLM, DeBERTa-MLM, BERT-SMM, and DeBERTa-SMM, each using the respective masking strategies discussed earlier. Table \ref{table:masking_techniques} summarizes the results across various combinations of models and masking methods.

\begin{table}[h]
\centering
\caption{Precision@20 Results for Different Masking Techniques Across Various Datasets}
\label{table:masking_techniques}
\begin{tabular}{lccc}
\hline
\textbf{Model} & \textbf{Diginetica} & \textbf{Yoochoose 1/64} & \textbf{Tmall} \\ \hline
DeBERTa (MLM) & 52.75 & 70.66 & 33.86 \\
BERT (MLM) & 52.98 & \underline{70.93} & 34.21 \\
GPT (CLM) & 52.45 & 70.42 & 33.40 \\
\hline
DeBERTa (SMM) & \underline{53.04} & 70.86 & \underline{36.08} \\ 
BERT (SMM) & \textbf{53.49} & \textbf{71.23} & \textbf{38.34} \\
\hline
\end{tabular}
\end{table}

As shown above, the \gls{smm} masking approach outperforms both CLM and \gls{mlm} techniques across all three datasets and the transformer architectures tested. While CLM, commonly used in auto-regressive models like GPT, predicts tokens sequentially based on prior context, and \gls{mlm}, employed in models like BERT, masks and predicts certain tokens based on surrounding context to capture bidirectional relationships, \gls{smm} is specifically designed for recommendation systems. It focuses on predicting the next item a user is likely to interact with, directly aligning with the goal of next-click prediction. This confirms the hypothesis \ref{hypothese1}.



\subsection{Sequentially Masking the Last K Items}\label{SMM-last-k}

In this section, we explain our decision to mask the penultimate token of each sub-sequence rather than the last one. Additionally, we explore whether masking the penultimate token yields better results compared to masking the third-to-last token. Table \ref{table:last_k_mask} presents the performance of the optimized BERT architecture under different masking strategies. The results indicate that, in some cases, masking the third-to-last token can produce better outcomes. However, this is highly dependent on the level of disorder in the dataset. In an ideal scenario with perfectly ordered data, masking only the last token would likely be the most effective. Unfortunately, such ideal conditions are rare in \gls{sbr} tasks, unlike in \gls{nlp}, where data typically follows a more fixed sequence.


Furthermore, the results confirm that masking the last token (\(K=1\)) consistently yielded the poorest performance, indicating that this approach is rarely optimal. Therefore, we use \textbf{\(K = 2\)} for all \gls{smm} training throughout the rest of this paper, as it provides a better balance between context utilization and accurate next-item prediction, confirming hypothesis \ref{hypothese3}.

\begin{table}[h]
\centering
\caption{Comparison of Different Last K Item Masked.}
\begin{tabular}{lccc}
\hline
\textbf{Model} & \textbf{Diginetica} & \textbf{Yoochoose 1/64} & \textbf{Tmall} \\ \hline
K = 1 & 52.76 & 70.48 & 35.16 \\
K = 2 & \textbf{53.49} & 71.23 & \textbf{38.42} \\
K = 3 & 53.47 & \textbf{71.24} & 38.40 \\
\label{table:last_k_mask}
\end{tabular}
\end{table}

\section{Comparison with the State of the Art}\label{SoA}

In this section, we compare our optimized encoder-only transformer architectures trained with the \gls{smm} method—against various state-of-the-art models. The results are presented in Table \ref{tab:results3.4.3}, which compares the \textit{P@20} and \textit{MRR@20} metrics across the Yoochoose 1/64, Diginetica, and Tmall datasets.

\subsection{Types of Benchmark Approaches}

In this section, we outline the different types of benchmark approaches used to compare against our models. These include single-session, cross-session, and multi-relation approaches, each varying in the amount of user session history and relational data they utilize for making recommendations \cite{zhang2023context}.


\textbf{Single-Session Approach}: This approach considers each session independently, without taking into account information from other sessions. It primarily relies on the content of the current session.

\textbf{Cross-Session Approach}: Cross-session approach incorporates information from multiple past sessions of a user to enhance the performance of recommendations. It leverages the history of sessions to understand long-term user preferences and behavior trends. This approach may use session merging techniques, where past sessions are aggregated or encoded to provide additional context during recommendations.

\textbf{Multi-Relation Approach}: This method considers various relationships between sessions and users to make more personalized and accurate recommendations. In other words, it uses information that goes beyond individual sessions, drawing from broader user-related data to generate recommendations.

Cross-session and multi-relation approaches benefit from access to more information during prediction, often leading to improved prediction performance. Since our prediction approach using the optimized transformer-SMM relies solely on single-session data, we focus on comparing its performance against state-of-the-art single-session approaches. Therefore, our performance analysis primarily benchmarks our method against other single-session models.

\begin{table*}[h]
    \centering
    \caption{Comparison of our models with state-of-the-art models. The best-performing model per approach is highlighted in \textbf{bold}, while the second-best model is underlined.}
    \label{tab:results3.4.3}
    \begin{tabular}{c c c c c c c}
        \hline
         & \multicolumn{2}{c}{\textbf{Diginetica}} & \multicolumn{2}{c}{\textbf{Tmall}} & \multicolumn{2}{c}{\textbf{Yoochoose 1/64}} \\
        \hline
        \textbf{Single-Session Approach} & \textbf{P@20} & \textbf{MRR@20} & \textbf{P@20} & \textbf{MRR@20} & \textbf{P@20} & \textbf{MRR@20} \\
        \hline
        POP & 1.18 & 0.28 & 2.00 & 0.90 & 6.71 & 0.58 \\
        Item-KNN & 35.75 & 11.57 & 9.15 & 3.31 & 51.60 & 21.81 \\
        FPMC & 22.14 & 6.66 & 7.32 & 2.01 & 45.62 & 15.01 \\
        GRU4Rec & 30.79 & 8.60 & 10.93 & 5.89 & 60.64 & 22.89 \\
        NARM & 48.32 & 16.00 & 23.00 & 10.64 & 68.32 & 28.63 \\
        STAMP & 46.62 & 15.13 & 20.47 & 9.36 & 68.74 & 29.67 \\
        SR-GNN & 50.73 & 17.59 & 27.57 & 13.57 & 70.57 & 30.94 \\
        LESSR & 51.71 & \underline{18.15} & 23.53 & 9.56 & 70.65 & 30.59 \\
        \textit{GPT-CLM} & 52.45 & 17.16 & 33.40 & 16.91 & 70.42 & 30.98 \\
        \textit{DeBERTa-MLM} & 52.75 & 17.79 & 33.86 & 17.23 & 70.66 & 31.61 \\    
        \textit{BERT-MLM} & \underline{52.98} & 17.81 & 34.21 & 17.96 & \underline{70.93} & \underline{31.86} \\
        \textit{DeBERTa-SMM} & 52.85 & 18.02 & \underline{36.08} & \underline{19.59} & 70.86 & 31.84 \\
        \textit{BERT-SMM} & \textbf{53.49} & \textbf{18.63} & \textbf{38.34} & \textbf{21.19} & \textbf{71.23} & \textbf{32.05} \\
        \hline
        \textbf{Cross-Session Approach} \\ 
        \hline
        CSRM & 48.49 & 17.13 & 29.46 & 13.96 & - & - \\
        CoSAN & 51.97 & 17.92 & 32.68 & 14.09 & - & - \\
        GCE-GNN & \underline{54.82} & 19.04 & 31.42 & 14.05 & 70.91 & 30.63 \\
        S$^2$-DHCM & 51.38 & 18.44 & 31.26 & 13.73 & \underline{71.88} & \underline{31.32} \\
        MTD & 51.82 & 17.26 & 30.41 & 13.15 & - & - \\
        COTREC & 54.18 & \underline{19.07} & \underline{36.35} & \underline{18.04} & - & - \\
        CARES & \textbf{56.49} & \textbf{23.22} & \textbf{38.77} & \textbf{18.37} & \textbf{72.21} & \textbf{34.40} \\
        \hline
        \textbf{Multi-Relation Approach} \\ 
        \hline
        AutoGSR & \textbf{54.56} & \textbf{19.16} & \underline{33.71} & \underline{15.87} & \textbf{71.77} & \textbf{31.02} \\
        MGIR & - & - & \textbf{36.41} & \textbf{17.42} & - & - \\
        \hline
    \end{tabular}
\end{table*}


\subsection{Benchmark Models}

To evaluate the performance of our models, we compared them with 17 other models, spanning single-session, cross-session, and multi-relation approaches. 


\textbf{Single-Session Approach}:
\begin{itemize}
    \item \textbf{POP} recommends the most popular items.
    \item \textbf{Item-KNN} \cite{sarwar2001item} recommends items based on the cosine similarity between the items in the current session and candidate items.
    \item \textbf{FPMC} \cite{rendle2010factorizing} uses both Markov chains and matrix factorization to incorporate personalized and general user information.
    \item \textbf{GRU4REC} \cite{hidasi2015session} leverages the memory of GRU by modeling the entire sequence.
    \item \textbf{NARM} \cite{li2017neural} and \textbf{STAMP} \cite{liu2018stamp} use the attention mechanism to capture the user's current and general interest.
    \item \textbf{SRGNN} \cite{xu2019graph} and \textbf{LESSER} \cite{chen2020handling} convert each session into a graph without using inter-session information.
\end{itemize}

\textbf{Cross-Session Approach}:
\begin{itemize}
    \item \textbf{CSRM} \cite{wang2019collaborative} incorporates relevant information from neighboring sessions via a memory network.
    \item \textbf{CoSAN} \cite{luo2020collaborative} uses a multi-head attention mechanism to build dynamic representations of items by merging item representations from collaborative sessions.
    \item \textbf{GCE-GNN} \cite{wang2020global} and \textbf{MTD} \cite{huang2021graph} simultaneously focus on both inter-session and intra-session dependencies.
    \item \textbf{COTREC} \cite{xia2021self} and \textbf{S2-DHCN} \cite{xia2021selfaaai} employ a global argumentative view of items to extract informative self-supervision signals.
    \item \textbf{CARES} \cite{zhang2023context} uses contextual information from multiple sessions to enhance recommendations, even during evaluation on the test set.
\end{itemize}

\textbf{Multi-Relation Approach}:
\begin{itemize}
    \item \textbf{AutoGSR} \cite{chen2022autogsr} and \textbf{MGIR} \cite{han2022multi} both learn multi-faceted item relations to improve session representation. Note that MGIR uses inter-session information, while AutoGSR does not.
\end{itemize}

\subsection{Global Results Table}\label{sec:tabl-des-result}

Table \ref{tab:results3.4.3} compares the performance of our transformer models against several state-of-the-art models across the Diginetica, Tmall, and Yoochoose 1/64 datasets. The results highlight the effectiveness of our models, particularly those trained using the \gls{smm} method. Notably, BERT-SMM consistently outperforms all other models in the single-session approach category across the three datasets. When comparing our models to state-of-the-art cross-session and multi-relation methods, BERT-SMM remains highly competitive.

DeBERTa-SMM also demonstrates competitive results, coming in just behind BERT-SMM in most cases. For example, on the Tmall dataset, DeBERTa-SMM achieves a Precision@20 of 36.08, even outperforming several cross-session and multi-relation models.

These results underscore the strong performance of our optimized transformer models, particularly BERT-SMM, even when compared to models that have access to more comprehensive session or user-level data. The ability of our models to outperform in the single-session category demonstrates the effectiveness of the \gls{smm} method. With further enhancements, such as incorporating user-level or multi-session data, we believe our models have the potential to rival and even surpass state-of-the-art cross-session and multi-relation approaches.



\section{Conclusion}

In this paper, we introduced Sequential Masked Modeling (\gls{smm}), a novel masking technique specifically designed for encoder-only transformer models, along with broader architectural improvements that were applied to both encoder- and decoder-based models. By using data augmentation through sliding windows and masking the penultimate token in augmented sequences, \gls{smm} significantly improved next-click prediction performance. In combination with the architectural enhancements, we demonstrated strong performance gains in several key metrics across three widely-used session-based recommendation datasets: Yoochoose 1/64, Diginetica, and Tmall.

Our experimental results showed that the proposed BERT-SMM and DeBERTa-SMM models consistently outperformed traditional single-session approaches and remained competitive with state-of-the-art cross-session and multi-relation methods, despite being limited to single-session data. These findings validate the effectiveness of the \gls{smm} technique in capturing sequential dependencies and improving recommendation performance in session-based environments.

Furthermore, the success of BERT-SMM and DeBERTa-SMM in the single-session category suggests that there is significant potential for enhancing performance even further by incorporating additional context, such as cross-session data or multi-relational information. Future work could explore how these models perform when extended to utilize user-level histories, or how \gls{smm} can be adapted to improve performance in other recommendation domains. Overall, this study provides strong evidence for the efficacy of encoder-only transformer architectures in session-based recommendation tasks when enhanced by \gls{smm}, offering a promising direction for future research in this area.


\bibliographystyle{ACM-Reference-Format}
\bibliography{sample-base}

\clearpage

\appendix

\section{Appendix}

\subsection{Ablation study}\label{ablation_study}

\begin{figure}[h]
\centering
\includegraphics[width=3.5in]{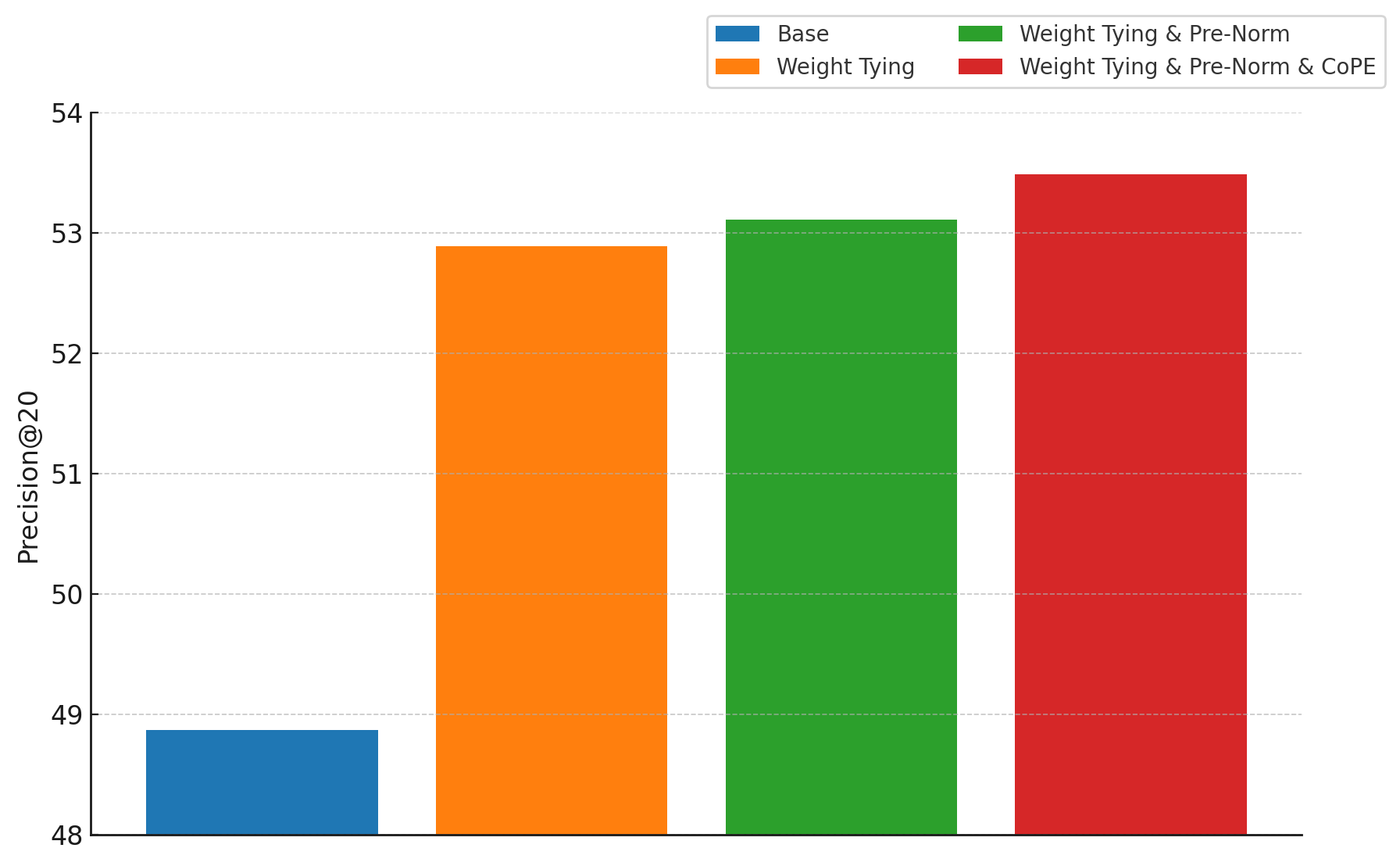}
\captionsetup{justification=centering}
\caption{Performance impact of various enhancement techniques applied to the BERT architecture trained with the Sequential Masked Modeling objective.}
\label{fig:Ablation}
\end{figure}

Our experiments demonstrate that the improved BERT architecture, combined with the \gls{smm} technique, delivers the best overall performance across the three datasets. In this section, we present an ablation study to assess the individual contribution of each optimization technique integrated into BERT-SMM. The results of this analysis, visualized in Figure \ref{fig:Ablation}, show how progressively adding these optimizations boosts performance on Diginetica.

\textbf{Different categories of ablation:}

\begin{itemize} 
    \item \textbf{Base}: This represents the standard BERT architecture as described in the original paper \cite{devlin2018bert}, trained with the SMM objective.
    
    \item \textbf{Weight Tying}: Reduces the number of parameters by sharing weights between input embeddings and the output layer.

    \item \textbf{Pre-layer Normalization}: Stabilizes training and improves convergence by applying normalization before attention layers.

    \item \textbf{Contextual Positional Encoding}: Enhances positional embeddings by incorporating contextual information instead of fixed token positions.

\end{itemize}

\subsection{Optimized BERT Architecture}\label{bert_optimized}

\begin{figure}[h]
\centering
\includegraphics[width=0.7\linewidth]{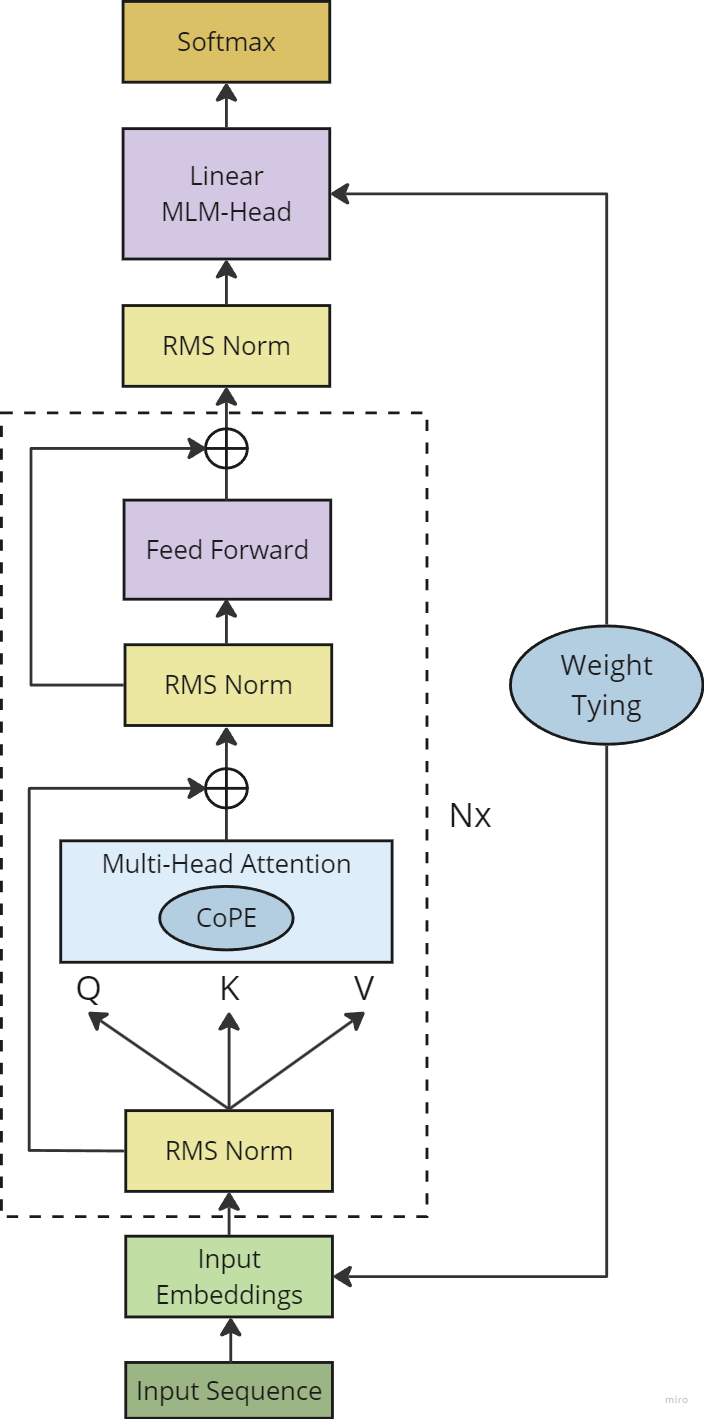}
\caption{Optimized BERT model architecture.}
\label{fig:bert_architecture}
\end{figure}

Figure \ref{fig:bert_architecture} shows the optimized BERT architecture that incorporates several key enhancements. These include pre-normalization with RMSNorm, Weight Tying, and Contextual Positional Encoding within the attention module. This design draws inspiration from modern transformer architectures, such as Llama \cite{touvron2023llama}, to boost performance and stability.

\subsection{Hyperparameter Setup}

Our model configuration is based on the BERT-Medium architecture from \textit{HuggingFace} \cite{huggingfaceBERTMedium}. The training setup was consistent across all experiments to ensure fair comparison. All hyperparameter values can be found in Table \ref{tab:hyperparams_transformer}, where we only conducted a search over the batch size. The model includes a maximum sequence length of 30, GeLU as the activation function, and a hidden size of 512. The optimizer used is Adam with an initial learning rate of 5e-5. Additionally, we applied a dynamic learning rate schedule that adjusts depending on the epoch, as detailed below.

The following learning rate scheduler was used: A custom learning rate schedule was applied using the function $\lambda_{\text{lr}} = 0.1$ if the epoch is 3 or higher, otherwise $\lambda_{\text{lr}} = 1.0$.

\begin{table}[H]
\centering
\caption{Hyperparameter configuration}
\label{tab:hyperparams_transformer}
\begin{tabular}{|l|l|}
\hline
\textbf{Hyperparameter}           & \textbf{Values}              \\ \hline
Batch Sizes                       & 32, 64, 128, 252             \\ \hline
Optimizer                         & Adam                         \\ \hline
Learning Rate                     & 5e-5                         \\ \hline
Model Max Length                  & 30                           \\ \hline
Activation Function               & GeLU                         \\ \hline
Hidden Size                    & 512                         \\ \hline
Number of Attention Heads         & 8                            \\ \hline
Depth (Layers)                    & 8                            \\ \hline

\end{tabular}
\end{table}

\end{document}